\documentclass[twocolumn]{pasj00}

\begin{document}
\SetRunningHead{K. Wajima et al.}{Milliarcsecond-Scale Structure in
PKS 1622$-$297}
\Received{2005/07/13}
\Accepted{2005/10/23}

\title{Milliarcsecond-Scale Structure \\ in the Gamma-Ray Loud Quasar
PKS 1622$-$297}


\author{%
Kiyoaki \textsc{Wajima}\altaffilmark{1},
Hayley E. \textsc{Bignall}\altaffilmark{2},
Hideyuki \textsc{Kobayashi}\altaffilmark{3,4},
Hisashi \textsc{Hirabayashi}\altaffilmark{5,6}, \\
Yasuhiro \textsc{Murata}\altaffilmark{5,6},
Philip G. \textsc{Edwards}\altaffilmark{5,6},
Masato \textsc{Tsuboi}\altaffilmark{4,7,8},
and
Kenta \textsc{Fujisawa}\altaffilmark{9}}

\altaffiltext{1}{Korea Astronomy and Space Science Institute, 61-1 Hwaam-dong,
Yuseong, Daejeon 305-348, Korea}
\email{kiyoaki@trao.re.kr}

\altaffiltext{2}{Joint Institute for VLBI in Europe, Postbus 2, NL-7990 AA
Dwingeloo, The Netherlands}

\altaffiltext{3}{National Astronomical Observatory of Japan, 2-21-1 Osawa,
Mitaka, Tokyo 181-8588}

\altaffiltext{4}{Department of Astronomy, School of Science, The University
of Tokyo, 7-3-1 Hongo, Bunkyo, Tokyo 113-0033}

\altaffiltext{5}{Institute of Space and Astronautical Science, Japan Aerospace
Exploration Agency, \\ 3-1-1 Yoshinodai, Sagamihara, Kanagawa 229-8510}

\altaffiltext{6}{Department of Space and Astronautical Science, The Graduate
University for Advanced Studies, \\ 3-1-1 Yoshinodai, Sagamihara, Kanagawa
229-8510}

\altaffiltext{7}{Nobeyama Radio Observatory, Minamimaki, Minamisaku, Nagano
384-1305}

\altaffiltext{8}{Department of Astronomical Science, The Graduate University
for Advanced Studies, \\ 2-21-1 Osawa, Mitaka, Tokyo 181-8588}

\altaffiltext{9}{Department of Physics, Faculty of Science, Yamaguchi
University, \\ 1667-1 Yoshida, Yamaguchi, Yamaguchi 753-8512}



\maketitle

\begin{abstract}

We have made a high-resolution VLBI observation of the gamma-ray loud quasar
PKS 1622$-$297 with the HALCA spacecraft and ground radio telescopes at
5\,GHz in 1998 February, almost three years after the source exhibited a
spectacular GeV gamma-ray flare.
The source shows an elongated structure toward the west on the parsec scale.
The visibility data are well modeled by three distinct components; 
a bright core and two weaker jet components.
Comparison with previous observations confirms that the jet components have
an apparent superluminal motion up to 12.1\,$h^{-1}c$, with the inner jet
components having lower superluminal speeds.
We apply the inverse Compton catastrophe model and derive a Doppler factor,
$\delta$, of 2.45, which is somewhat lower than that of other gamma-ray loud
active galactic nuclei (AGNs), suggesting the source was in a more quiescent
phase at the epoch of our observation.
As an alternative probe of the sub-parsec scale structure, we also present
the results from multi-epoch ATCA total flux monitoring, which indicate the
presence of persistent intraday variability consistent with refractive
interstellar scintillation.
We examine the gamma-ray emission mechanism in the light of these
observations.

\end{abstract}

\section{Introduction}
\label{sec:Introduction}

Observations made with the EGRET (Energetic Gamma Ray Experiment Telescope)
gamma-ray detector onboard the CGRO (Compton Gamma Ray Observatory) spacecraft 
resulted in the identification of nearly 70 gamma-ray emitting AGNs
\citep{Hartman99}, with more recent studies increasing the number of EGRET
sources with at least one plausible AGN association by $\sim 50$
(\cite{Sowards03}, \yearcite{Sowards04}).
These are predominantly radio-loud AGNs, which suggests a strong connection
between the gamma-ray emission and the radio emission.

Recent VLBI observations have revealed several interesting features of 
the parsec-scale structures in gamma-ray loud AGNs.
\citet{Jorstad01a} completed multi-epoch VLBA (Very Long Baseline Array)
observations of 42 gamma-ray loud AGNs at 22.2 and 43.2\,GHz, and 
occasionally 8.4 and 15.4\,GHz.
From these they found that 10 gamma-ray flares in eight objects fall within
1\,$\sigma$ uncertainties of extrapolated epoch of zero separation from
the core of a superluminal radio component.
In addition they compared their observations with variability of polarized
radio flux and gamma-ray flares, and suggest that gamma-ray emission occurs
in the superluminal radio components \citep{Jorstad01b}.
The fact that the gamma-ray flare occurs close to the time of the maximum
polarized radio flux density and after the mm-wavelength radio flare favors
a model in which the gamma-ray emission is produced by inverse Compton
scattering by the electrons in the parsec-scale jet rather than closer to
the central engine.

Gamma-ray emission from inverse Compton scattering (e.g. \cite{Sikora94})
requires a smaller jet viewing angle and higher Doppler factor because the
emission should be concentrated in the jet direction.
\citet{Jiang98} compiled the published data for 52 AGNs (including 18 EGRET
AGNs), for which measurements of the angular size and radio flux density of
the VLBI core, proper motion of the components in the jet, and X-ray flux
density were available.
They applied the inhomogeneous jet model of \citet{Konigl81} to the sample
and showed that the derived mean viewing angle for 14 EGRET sources is
\timeform{4D.9} and the Doppler factors of these sources are greater than 4.7.
These results also indirectly support the gamma-ray emission model by
inverse Compton scattering.
However, there are still large uncertainties for the derived parameters in
gamma-ray loud AGNs and it is therefore important to verify the milliarcsecond
(mas)-scale properties for individual sources precisely by VLBI observations.

In this paper we report the results of an observation of PKS 1622$-$297
(J1626$-$2951) by VSOP (VLBI Space Observatory Programme), consisting of
the radio-astronomy spacecraft HALCA and ground radio telescopes located
around the world \citep{Hirabayashi98}.
VSOP observations enable improvement of the angular resolution by up to
a factor of three compared to ground-only VLBI observations at the same
frequencies.

PKS 1622$-$297 is one of the gamma-ray emitting sources with a high-confidence
identification by EGRET \citep{Hartman99}.
During an outburst at epoch 1995.48, the source was the most luminous 
gamma-ray emitting AGN ever detected and showed rapid gamma-ray variability
by a factor of at least 3.6 in less than 7.1 hours \citep{Mattox97}.
It is a strong, compact radio source optically identified as a quasar
\citep{Saikia87,Stickel94} with a fractional polarization of more than 4\%
at cm-wavelengths \citep{Tabara80,Impey90}.
Previous results of VLBI observations are described by \citet{Tingay98},
\citet{Fomalont00}, \citet{Jorstad01a}, and \citet{Tingay02}.

PKS 1622$-$297 has also been reported to display intraday variability (IDV)
at 8.6\,GHz at two epochs in 1994 \citep{Kedziora01}.
Both intrinsic and extrinsic origins have been suggested for IDV
\citep{Wagner95}, while current radio observations show that interstellar
scintillation (ISS) is mainly responsible for the rapid IDV feature, such
as the annual modulation in the scintillation timescale
\citep{Rickett01,Jauncey01,Bignall03b}, and detection of time delay across
the 10,000\,km baseline between ATCA and VLA \citep{Jauncey03}.
Both intrinsic and extrinsic explanations require very compact source
structure.
We present here the results of multi-frequency multi-epoch monitoring of
PKS 1622$-$297 with the Australia Telescope Compact Array (ATCA) and
discuss the origin of the IDV, and consider the gamma-ray emission
mechanism based on both the VSOP and ATCA observations.

A redshift of $z = 0.815$ was listed in \citet{Wright90}, corresponding
to an angular-to-linear scale conversion of 5.3\,$h^{-1}$\,pc\,mas$^{-1}$
assuming a Hubble constant, $H_0$, of 100\,$h$\,km\,s$^{-1}$\,Mpc$^{-1}$,
a flat universe with a cosmological constant, $\Omega_{\Lambda} = 0.7$,
and a pressureless matter density parameter, $\Omega_m = 0.3$
\citep{Balbi00}.
However, this redshift has been questioned \citep{Jackson02}, and so
although we adopt it in this paper, it should be kept in mind that
confirmation of this value is required.

The details of the observations and data reduction are described and the
results are presented in section \ref{sec:VSOP Observation} and
\ref{sec:ATCA Observations} for VSOP and ATCA observations, respectively.
In section \ref{sec:Discussion}, we discuss the mas-scale structure and the
gamma-ray emission mechanism in the light of these observations.
Throughout this paper, we define the spectral index, $\alpha$, as
$S_{\nu} \propto \nu^{\alpha}$.


\section{VSOP Observation}
\label{sec:VSOP Observation}

\subsection{Observation and Data Reduction}

The VSOP observation was made in left-circular polarization at 5\,GHz on
1998 February 22, as a part of the VSOP AGN Survey Program
\citep{Hirabayashi00}.
The ground radio telescopes were Hartebeesthoek 26\,m (South Africa), Hobart
26\,m (Australia), and Shanghai 25\,m (China).
Data from HALCA were transferred via the Tidbinbilla (Australia) tracking
station.
The data were correlated at the Dominion Radio Astrophysical Observatory
S2 correlator \citep{Carlson99} in Penticton (Canada).
Output pre-averaging time were 0.1 second for space--ground baselines and
2 seconds for ground--ground baselines, respectively.
The observation bandwidth was 32\,MHz and the total observing time was
4.5 hours.

We carried out post-processing of the correlated data using the Astronomical
Image Processing System (AIPS) software developed by the National Radio
Astronomy Observatory (NRAO) and the Caltech software Difmap
\citep{Shepherd97}.
We applied {\it a priori} amplitude calibration using the antenna gain
factors and system noise temperature measurements.
For fringe-fitting, we used a solution interval of three minutes and
a point-source model, with Shanghai serving as the reference telescope.
We detected fringes on all ground--ground baselines.
On the space--ground baselines, however, we could detect significant fringes
only within an hour from the beginning of the observation, corresponding
to the shorter space baselines, and so we excluded the HALCA data after
that time.

\begin{figure}
\begin{center}
\FigureFile(80mm,82mm){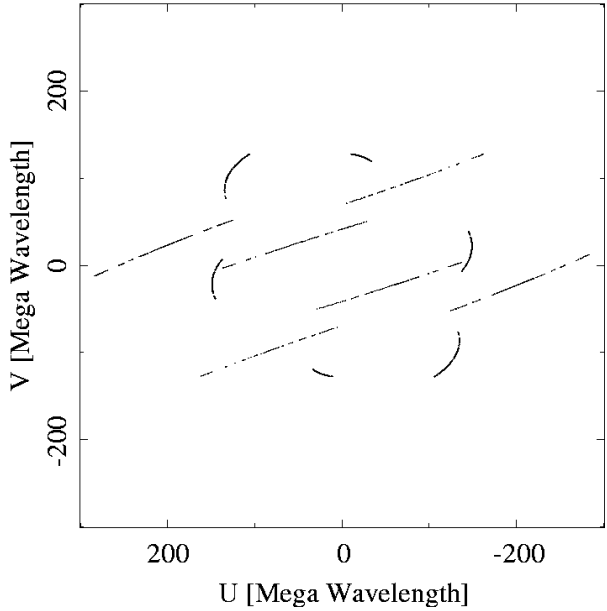}
\end{center}
\caption{($u,v$) coverage for the VSOP observation of PKS 1622$-$297 at
epoch 1998.15 (1998 February 22).}
\label{fig:Figure1}
\end{figure}

We exported the fringe-fitted visibility data to Difmap for imaging.
At first we integrated the data over 10 seconds to reconcile the different
preaveraging time from the correlator output, and performed several iterations
of cleaning and self-calibration to the phases.
To ensure a better angular resolution with the HALCA data, we adopted uniform
weighting of the data with gridding weights scaled by amplitude errors
raised to the power of $-1$ \citep{Hirabayashi00}.
Figure~\ref{fig:Figure1} shows the ($u,v$) coverage of all baseline
pairs for the times that fringes were successfully detected.
The elliptical tracks represent the baselines between ground stations, while
space-ground baselines contribute to the 
approximately linear inner and outer tracks.
The longest baseline of the observation is 17,000\,km, corresponding to
1.3 times the Earth's diameter.


\subsection{VSOP Results}

\begin{figure}
\begin{center}
\FigureFile(80mm,82mm){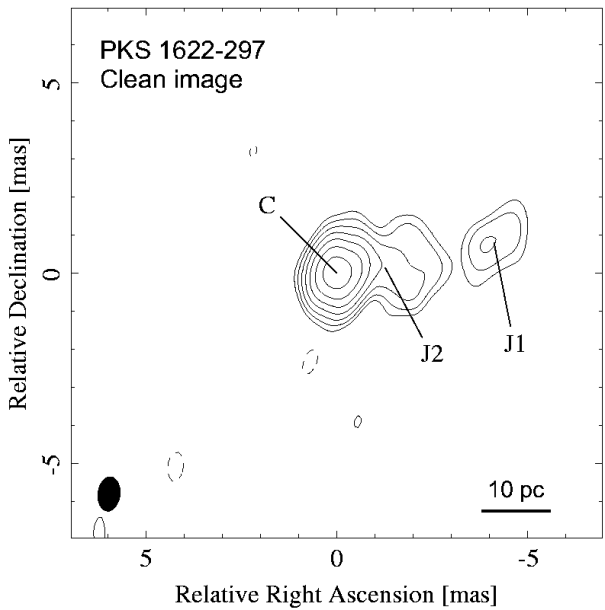}
\end{center}
\caption{5\,GHz VSOP image of PKS 1622$-$297 at epoch 1998.15 (1998 February
22). The restoring beam is 0.91\,mas $\times$ 0.58\,mas at a position angle
of $\timeform{7D}$, which is indicated at the lower left corner. Contour
levels are 9.5 $\times$ ($-1$, 1, 2, 4, ..., and 64)\,mJy\,beam$^{-1}$, and
the peak flux density is 0.84\,Jy\,beam$^{-1}$. The labels show the Gaussian
model fitting components. See table \ref{tbl:Table1} for detailed
parameters.}
\label{fig:Figure2}
\end{figure}

Figure~\ref{fig:Figure2} shows the image of PKS 1622$-$297.
We detect three components, which are the bright core, C, a moderately bright
jet component, J2, 1.3\,mas from the core, and a weak jet component, J1,
4.2\,mas from the core.
The jet extends to the west, which is consistent with previous results
(Tingay et~al.\ 1998; \cite{Fomalont00}; \cite{Jorstad01a}; \cite{Tingay02}).
However we did not detect the stationary component $\sim 15$\,mas from the
core, which was detected by \citet{Tingay98} and \citet{Fomalont00} at
5\,GHz, and \citet{Jorstad01a} at 15\,GHz.
This component is relatively faint in the 5\,GHz observation
(Tingay et~al. 1998) and even weaker ($\sim 30$\,mJy) in the 15\,GHz
observations \citep{Jorstad01a}.
The flux density of the component at 5\,GHz was 50\,mJy at epoch 1994.64
(Tingay et~al. 1998) and 40\,mJy at epoch 1996.43 \citep{Fomalont00}.
The upper limit of the component on our image at epoch 1998.15 is 20\,mJy,
which corresponds to the 7\,$\sigma$ detection limit.
Although it is likely that the component is resolved, the sparse ($u,v$)
coverage of our observation may result in a dynamic range that is not
adequate to detect the component.

\begin{figure}
\begin{center}
\FigureFile(80mm,60mm){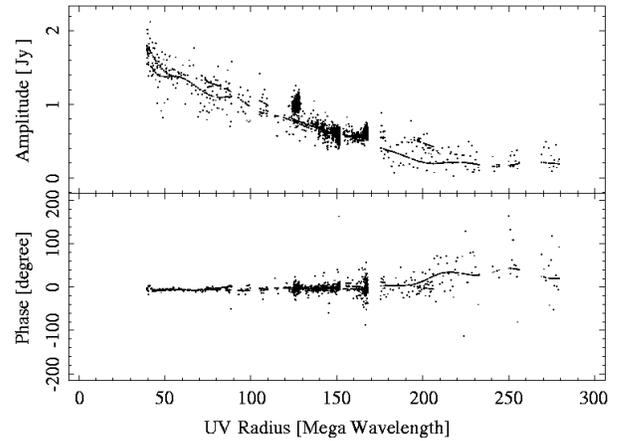}
\end{center}
\caption{Calibrated visibility amplitudes (top) in Jy and phases (bottom)
in degree as a function of the ($u,v$) distance. Dotted lines show the clean
component model used for restoring the image in figure
\ref{fig:Figure2}.}
\label{fig:Figure3}
\end{figure}

Figure~\ref{fig:Figure3} shows the calibrated visibility amplitudes and
phases as a function of ($u,v$) distance. 
The clustered visibilities in the range between 120\,M$\lambda$ and
170\,M$\lambda$ are the ground-ground baseline data, while the space-ground
visibilities are widely distributed up to 280\,M$\lambda$.
The space-ground visibilities show that the source is clearly resolved.
We could not detect significant fringes on baselines longer than
280\,M$\lambda$, indicating that the source is being resolved out on those
baselines.

The shortest baselines sampled in the observation are $\sim 40$\,M$\lambda$,
as shown in figure~\ref{fig:Figure3}.
PKS 1622$-$297 was fortuitously observed one week earlier as part of the
multi-epoch ATCA monitoring of VSOP Survey Program sources, yielding flux
densities on the ATCA 6\,km baseline of 2.33\,Jy at 1.384\,GHz, 2.23\,Jy at
2.496\,GHz, 2.41\,Jy at 4.800\,GHz, and 2.49\,Jy at 8.640\,GHz
\citep{Tingay03}.
A single-dish measurement from the University of Michigan Radio Astronomy
Observatory (UMRAO) gave an 8.0\,GHz flux density of the source of
$2.49 \pm 0.06$\,Jy on 1998 February 28, in excellent  agreement with the
ATCA value.
The spectral index between 4.8 and 8.6\,GHz at this epoch was comparatively
flat, $\alpha \sim 0.1$:
Figure~2 of \citet{Tingay03} reveals that the spectrum was more inverted
earlier and later in the monitoring program.
The contemporaneous flux density measurement is consistent with a simple
extrapolation of our correlated flux density to smaller baselines.
We also examined our amplitude calibration using {\tt gscale} in Difmap
and resulted in correction factors of less than 10\% for all stations.
We therefore believe the amplitude calibration of our observation to be
reliable.


\subsection{Model Fitting and Proper Motion of Each Component}

To quantify the size and relative location of components, we modeled the
calibrated visibility data with elliptical Gaussian components.
Table~\ref{tbl:Table1} gives the model fitting results and the position of
each component is indicated in figure~\ref{fig:Figure2}.
\begin{table*}
\begin{center}
\caption{Model fitting results.}
\label{tbl:Table1}
\begin{tabular}{cccccccc} \hline\hline
Component & $S$\footnotemark[$*$]          &
$r$\footnotemark[$\dagger$]                &
$\phi$\footnotemark[$\ddagger$]            &
$\theta_{\mathrm{maj}}$\footnotemark[$\S$] &
$\theta_{\mathrm{min}}$\footnotemark[$\|$] &
P.A.\footnotemark[$\#$]                    &
$T_{\mathrm{B}}$\footnotemark[$**$]        \\
 & (Jy) & (mas) & (deg) & (mas) & (mas) & (deg) & ($10^{11}$ K) \\
\hline
C  & $1.50 \pm 0.01$ & 0               & 0               & $0.71 \pm 0.01$
 & $0.54 \pm 0.01$ & $-78.9 \pm 1.7$ & $3.73 \pm 0.07$ \\
J2 & $0.34 \pm 0.02$ & $1.31 \pm 0.05$ & $-81.2 \pm 2.3$ & $2.09 \pm 0.08$
 & $1.05 \pm 0.07$ & $ ~~69.1 \pm 2.8$ & $0.15 \pm 0.01$ \\
J1 & $0.13 \pm 0.01$ & $4.28 \pm 0.07$ & $-78.1 \pm 0.9$ & $1.58 \pm 0.12$
 & $0.69 \pm 0.12$ & $-50.7 \pm 5.1$ & $0.11 \pm 0.02$ \\
\hline
\multicolumn{8}{@{}l@{}}{\hbox to 0pt{\parbox{170mm}{\footnotesize
Notes.
\par\noindent
\footnotemark[$*$] Flux density of each component.
\par\noindent
\footnotemark[$\dagger$] Distance of each component from the origin defined
by component C.
\par\noindent
\footnotemark[$\ddagger$] Position angle of each component with respect to
the origin (east of north).
\par\noindent
\footnotemark[$\S$], \footnotemark[$\|$], \footnotemark[$\#$] Parameters of
Gaussian model: major and minor axes in mas (FWHM) and the position angle
(P.A.) of the major axis.
\par\noindent
\footnotemark[$**$] Peak brightness temperature given in the rest frame of
the host galaxy (assuming $z = 0.815$; see section \ref{sec:Introduction}).
}\hss}}
\end{tabular}
\end{center}
\end{table*}
We performed the model fitting using the task `IMFIT' in AIPS.

\begin{figure*}
\begin{center}
\FigureFile(150mm,135mm){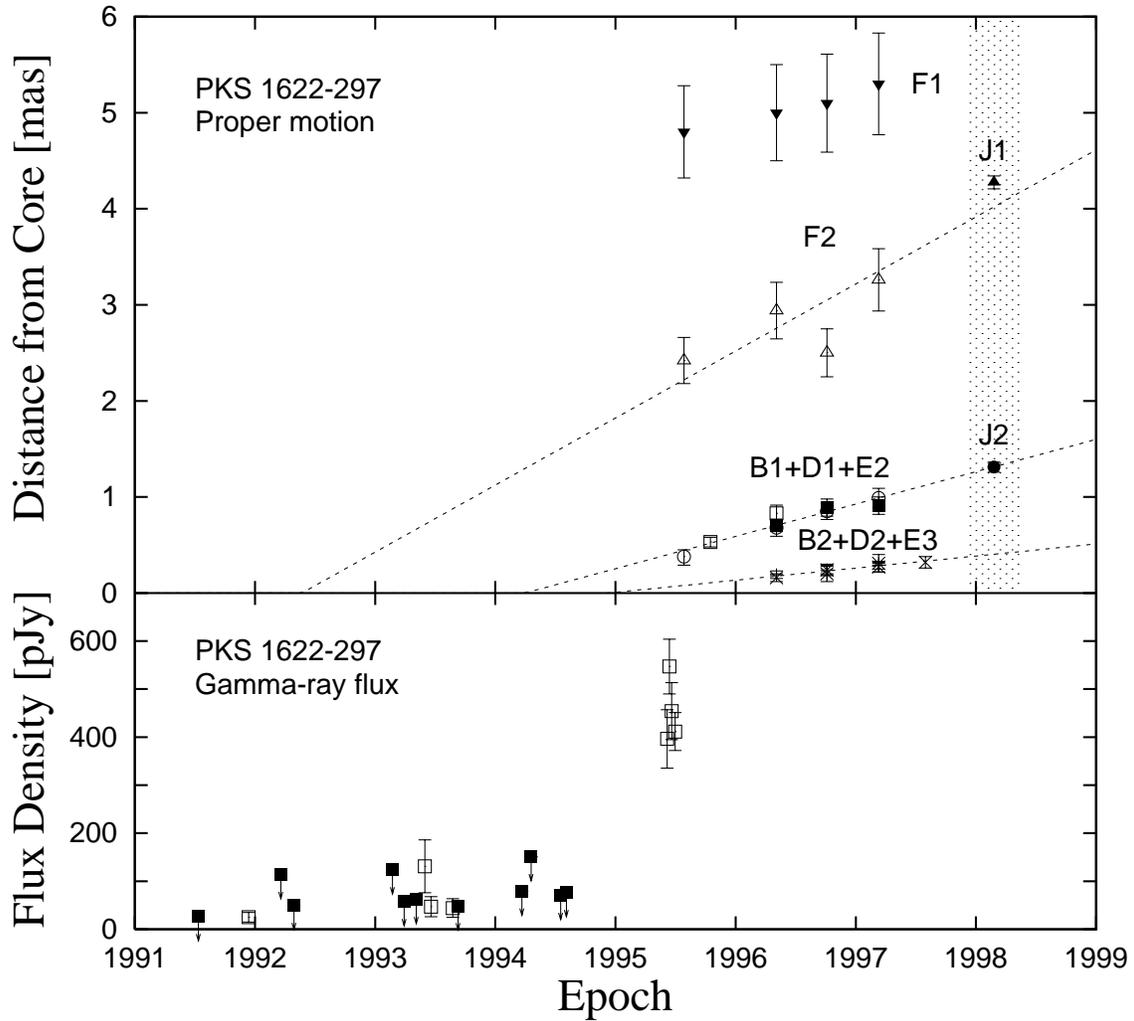}
\end{center}
\caption{(Top) Distance of components from the core as a function of observed
epoch. The measurements in 1998.15 (J1 and J2) are our results. See table
\ref{tbl:Table2} for numerical values. The data before 1997.58
(B2+D2+E3, B1+D1+E2, F2, and F1) are from \citet{Jorstad01a}. The dashed
lines show results of a least-squares linear fit. Position uncertainties
are not listed for some data points in \citet{Jorstad01a}, and we therefore
assume the uncertainties as 10\% of listed values.
(Bottom) Gamma-ray light curve in picoJy observed by EGRET \citep{Hartman99}.
Open and filled squares show the positive and marginal (or non-) detections,
respectively. Upper limits are plotted in the latter case.}
\label{fig:Figure4}
\end{figure*}

The data were modeled satisfactorily by three distinct components.
The brightest component, labeled C, is the brightest and most compact,
consistent with it being the core.
Weaker components, labeled J2 and J1, are situated to the west of the core.
We compared the positions of each component with those measured by
\citet{Jorstad01a}.
Figure~\ref{fig:Figure4} shows the distance of each component from
the core as a function of observed epoch.
The results of a least-squares linear fit for each component are also
plotted as dashed lines on figure~\ref{fig:Figure4}.
Component J2 can be identified with the B1+D1+E2 of \citet{Jorstad01a},
and J1 with their component F2.
\citet{Tingay02} have obtained an image of PKS 1622$-$297 with VSOP at a
different epoch from ours and detected a weak jet component toward the west
approximately 1.5\,mas from the core.
Although they do not apply model fitting to this component, they mention
that it is likely to be component B1+D1+E2.
Our results clearly show that these two components are identical.

\citet{Jorstad01a} have also detected the component F1, as shown in
figure~\ref{fig:Figure4}. F1's position was less well-defined, and its
motion appeared to be much sower than that of F2.
Extrapolating the motion of F1 suggests it would be located
$5.6 \pm 0.5$\,mas from the core at the epoch of our VSOP observation.
This position is slightly beyond the northwestern extension of J1 (see
figure~\ref{fig:Figure2}), and so it is possible that J1 may be a blended
component of F2 and F1.

We could not detect the component that corresponds to B2+D2+E3 detected by
\citet{Jorstad01a}.
If we assume a linear motion of B2+D2+E3, we would detect the component with
$\sim 0.4$\,mas separation from the core.
While there is emission evident at this location in figure~\ref{fig:Figure2},
it is adequately represented by our modelfits for components C and J2.
Our non-detection may be due to insufficient angular resolution.
However, we point out that \citet{Tingay02} also did not detect this component
in their VSOP image at epoch 2000.32.
Their observation had better angular resolution than ours and suggests
that B2+D2+E3 has either faded away or has a significantly different
speed than that determined by \citet{Jorstad01a}.

Apparent proper motions of $\mu = 0.13 \pm 0.01$ (for B2+D2+E3),
$0.34 \pm 0.02$ (for J2), and $0.70 \pm 0.11$ (for J1) correspond to
apparent velocities of $(2.2 \pm 0.2)$\,$h^{-1}$,
$(5.9 \pm 0.3)$\,$h^{-1}$, and $(12.1 \pm 1.9)$\,$h^{-1}$, respectively
(see also table~\ref{tbl:Table2}).
\begin{table*}
\begin{center}
\caption{Apparent proper motion of each component.}
\label{tbl:Table2}
\begin{tabular}{cccc}
\hline\hline
Component & Proper motion   & Apparent velocity & Ejection epoch \\
          & (mas yr$^{-1}$) &                   & (yr)           \\
\hline
B2+D2+E3     & $0.13 \pm 0.01$   & $(2.2 \pm 0.2)$\,$h^{-1}$\,$c$
 & $1994.94 \pm 0.08$ \\
B1+D1+E2, J2 & $0.34 \pm 0.02$   & $(5.9 \pm 0.3)$\,$h^{-1}$\,$c$
 & $1994.25 \pm 0.10$ \\
F2, J1       & $0.70 \pm 0.11$   & $(12.1 \pm 1.9)$\,$h^{-1}$\,$c$
 & $1992.39 \pm 0.73$ \\
\hline
\end{tabular}
\end{center}
\end{table*}
The ejection epochs listed in table~\ref{tbl:Table2} are derived by
extrapolating the linear relations of each component.
The ejection epoch of J1 is estimated to be $1992.39 \pm 0.73$.
This is coincident with a local maximum in the gamma-ray emission at
1991.95 although there is a relatively large error.
Although the gamma-ray light curve is under-sampled, the coincidence
suggests a close relationship between gamma-ray emission and ejection
of a new jet component.

\citet{Jorstad01a} pointed out that inner components have superluminal
speeds but they are lower than those of outer components.
The apparent speed of J2 is half that of J1 and our result,
combining with results by \citet{Jorstad01a}, confirms this tendency.
We will discuss features of each component derived from $\beta_{\mathrm{app}}$
and the Doppler factor in section \ref{subsec:Estimation of Parameters}.



\section{ATCA Observations}
\label{sec:ATCA Observations}

\subsection{Observation and Data Reduction}

PKS 1622$-$297 was included in a sample of 118 compact, flat-spectrum AGN
observed at two-epochs and four frequencies with ATCA as part of survey for
sources displaying IDV \citep{Kedziora01}.
Twenty-two sources (19\% of the sample) were found to display significant
IDV at at least one frequency and at least one epoch, with significant IDV
observed for PKS 1622$-$297 at 8.6\,GHz at both epochs.
This led to the inclusion of PKS 1622$-$297 in the sample of 21 sources
monitored with the ATCA by \citet{Bignall03a}.

PKS 1622$-$297 was monitored at 4.8\,GHz and 8.64\,GHz at ten epochs
between 2001 February 4 and 2002 February 21.
The observing technique and data reduction methods are described in detail
in \citet{Bignall03b}; briefly, observations at each epoch were generally
made over a two-day period.
Sources were observed in scans of up to 5 minutes every $\sim 1$ hour while
they were above an elevation of \timeform{15D}.
A number of calibrator sources were included in the observing schedule and
observed in a similar manner.

The data were reduced in MIRIAD \citep{Sault95}.
Time-dependent antenna gain corrections were derived at each frequency from
observations of the calibrators, and found to vary by $\lesssim 1$\% over
each epoch.
After applying these gain corrections, averaged over 30--60 minutes and
linearly interpolated, phase self-calibration was performed using a point
source model and 10 second integrations.
The real parts of the visibilities for Stokes parameters $I$, $Q$, and $U$
were averaged over intervals of several minutes.


\subsection{ATCA Results}

\begin{figure*}
\begin{center}
\FigureFile(84mm,60mm){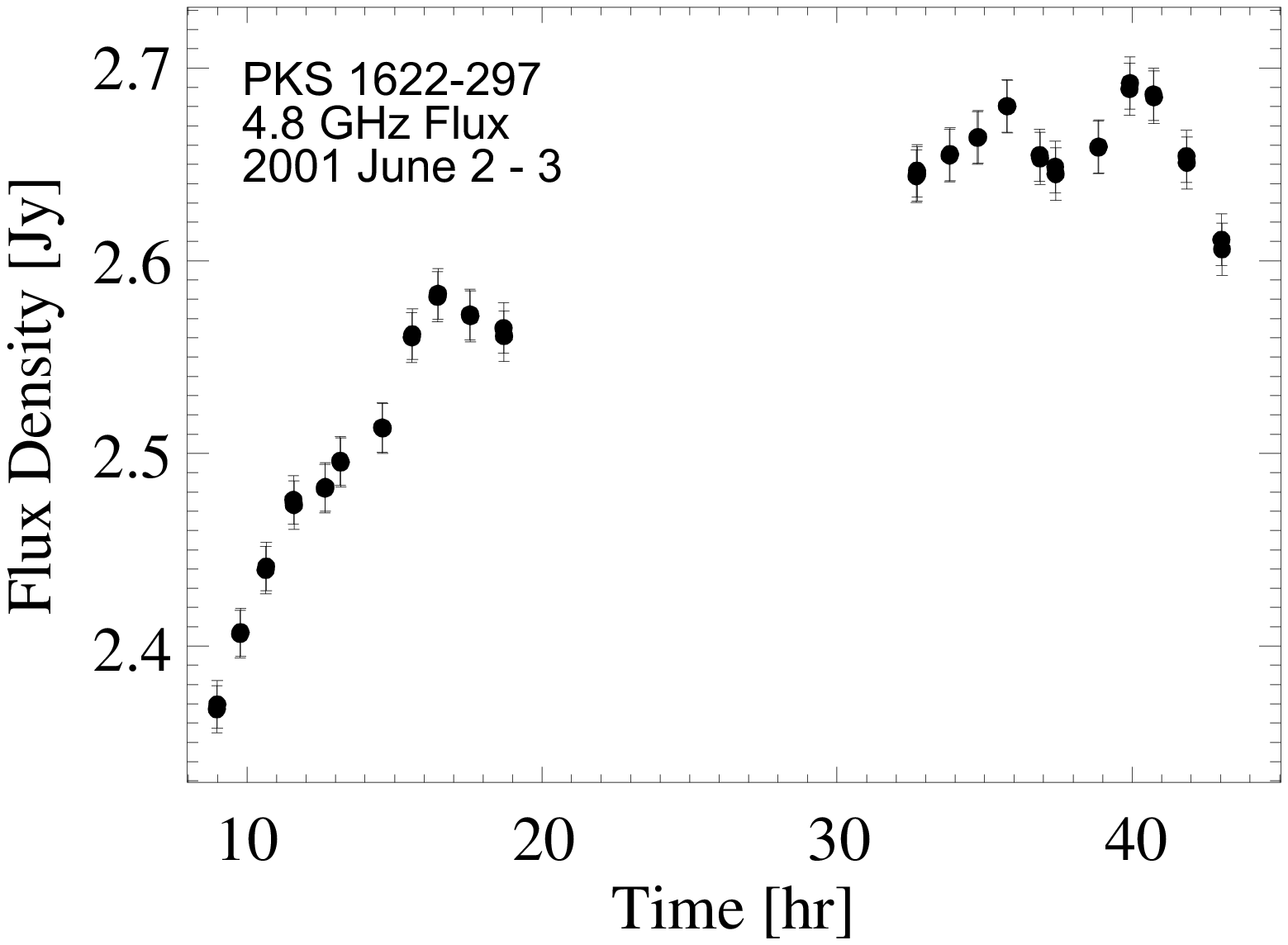}
\FigureFile(84mm,60mm){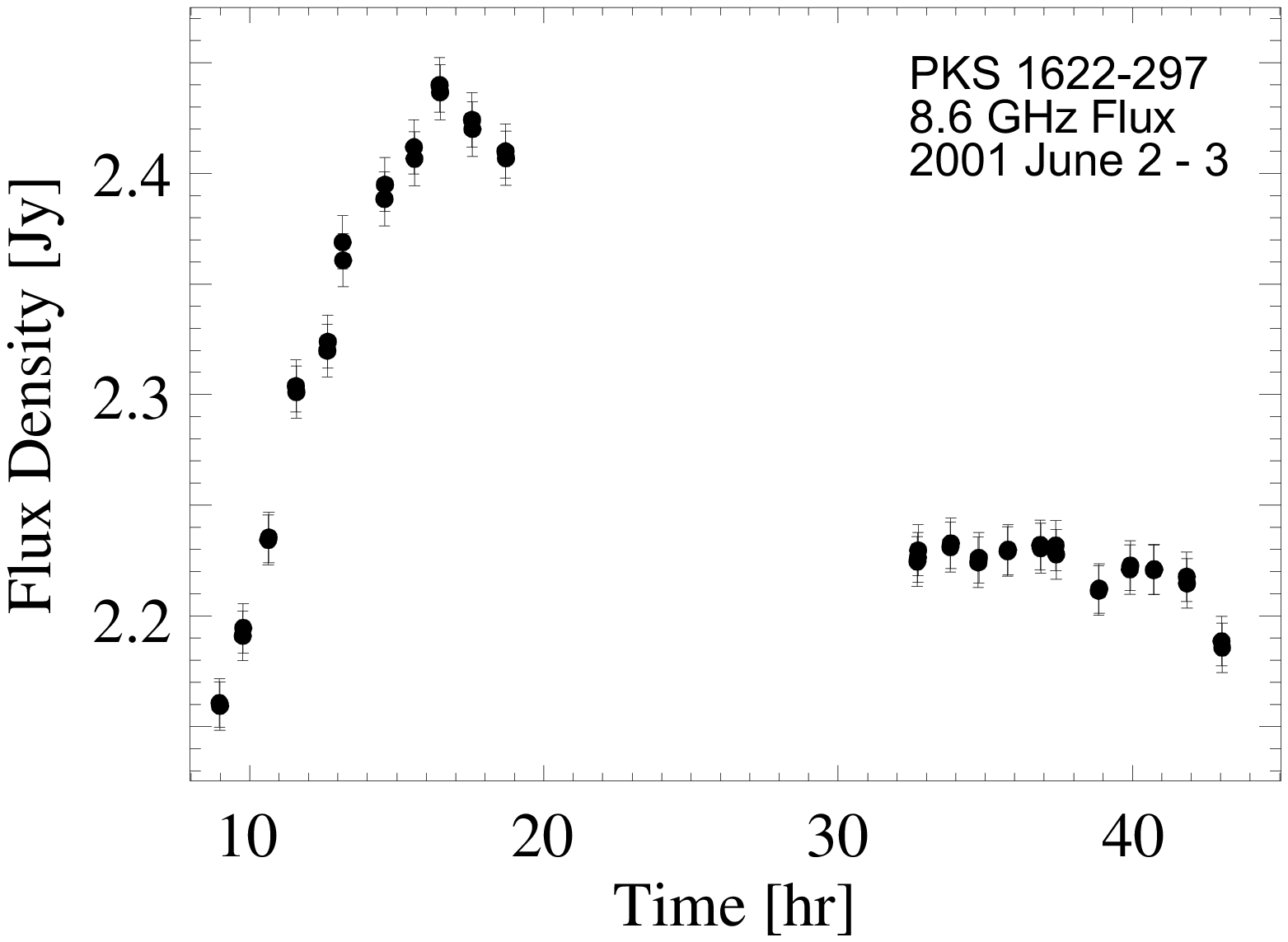}
\FigureFile(84mm,60mm){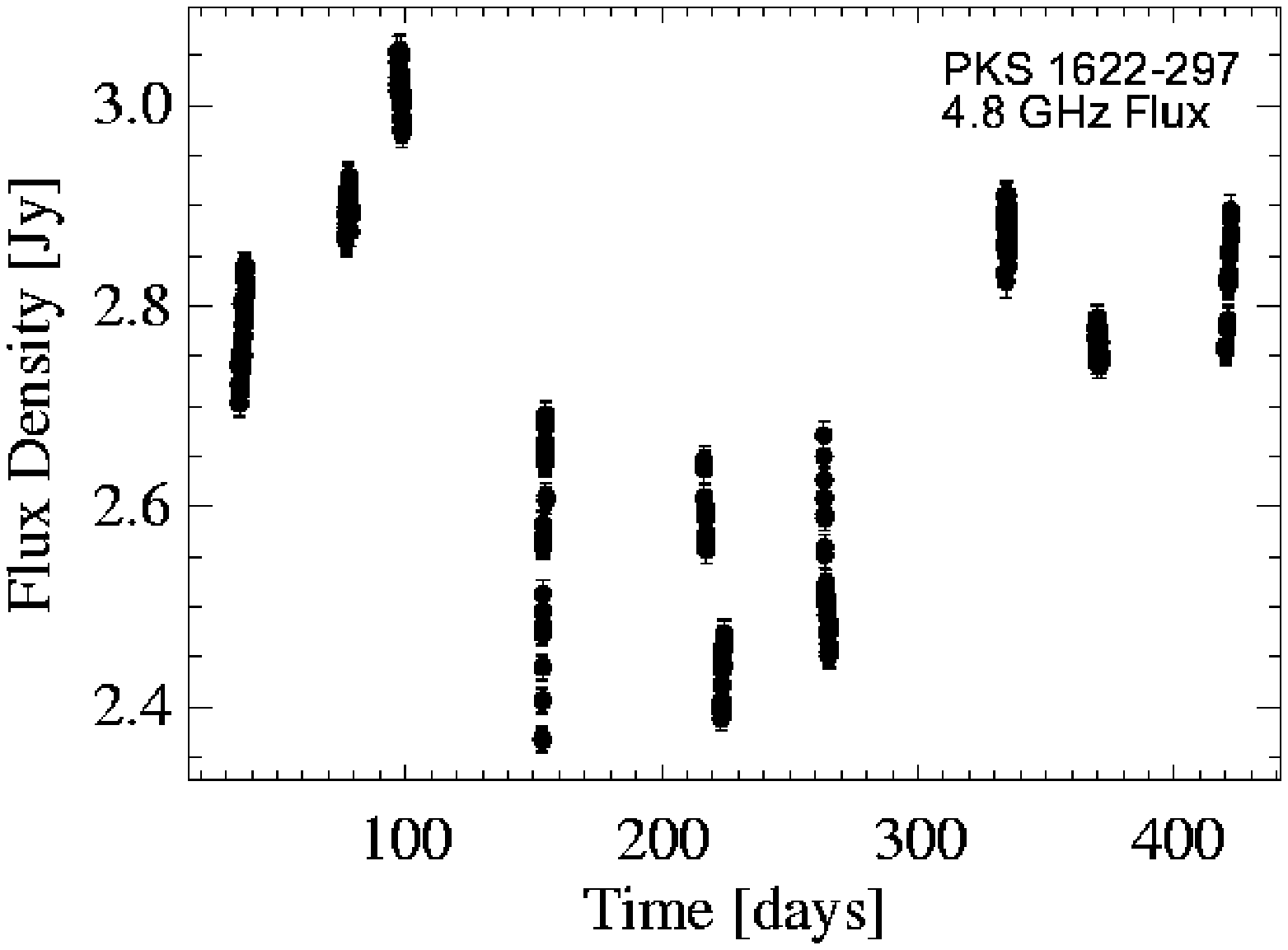}
\FigureFile(84mm,60mm){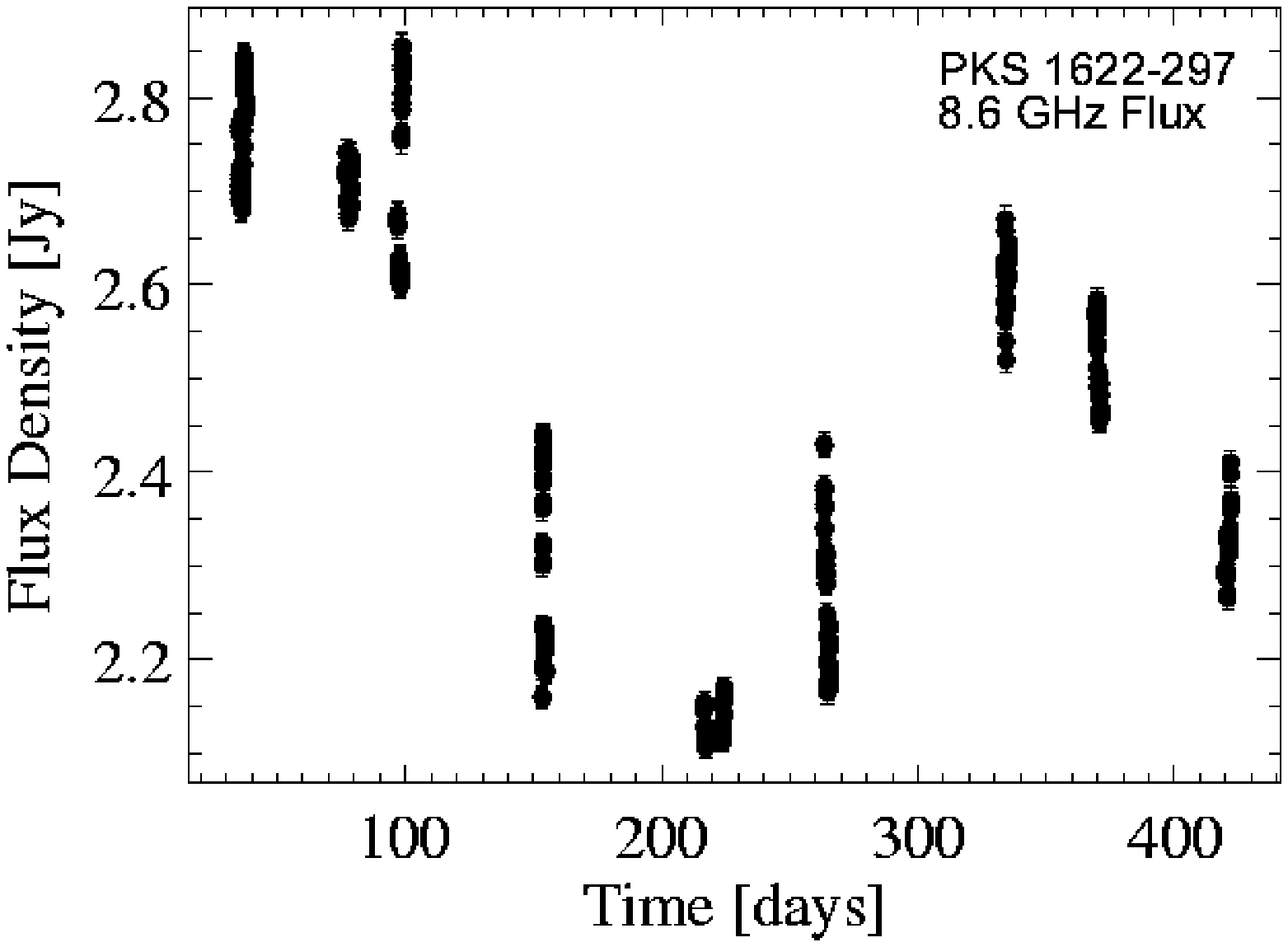}
\end{center}
\caption{Results of the flux density monitoring in the ATCA IDV survey
program at 4.8 and 8.6\,GHz. (Top) Total flux density at 4.8\,GHz (left)
and 8.6\,GHz (right) on 2001 June 2 -- 3, when the largest amplitude IDV was
observed. Horizontal axis shows hours UT since 2001 June 2 (day 153), 00:00.
(Bottom) Long-term flux variation at 4.8\,GHz (left) and 8.6\,GHz (right).
Horizontal axis shows days since 2001 January 1.}
\label{fig:Figure5}
\end{figure*}

The results of the ATCA monitoring are given in table~\ref{tbl:Table3}.
\begin{table*}
\begin{center}
\caption{ATCA monitoring results.}
\label{tbl:Table3}
\begin{tabular}{cccccccccc} \hline\hline
Epoch                                                     &
Time span                                                 &
$N_{4.8}$                                                 &
$\bar{S}_{4.8}$                                           &
$m_{4.8}$\footnotemark[$*$]                               &
$S_{\mathrm{max}}^{4.8} - S_{\mathrm{min}}^{4.8}$         &
$N_{8.6}$                                                 &
$\bar{S}_{8.6}$                                           &
$m_{8.6}$\footnotemark[$\dagger$]                         &
$S_{\mathrm{max}}^{8.6} - S_{\mathrm{min}}^{8.6}$         \\
 & (days) &  & (Jy) & (\%) & (Jy) &  & (Jy) & (\%) & (Jy) \\
\hline
2001 Feb  4 & 2.77 & 24 & 2.78 & 1.4 & 0.12 & 25 & 2.76 & 1.8 & 0.15 \\
2001 Mar 17 & 1.96 & 26 & 2.90 & 0.6 & 0.06 & 26 & 2.71 & 0.8 & 0.07 \\
2001 Apr  6 & 1.96 & 23 & 3.02 & 0.9 & 0.08 & 23 & 2.70 & 3.7 & 0.25 \\
2001 Jun  2 & 1.42 & 22 & 2.58 & 3.8 & 0.32 & 22 & 2.28 & 3.9 & 0.28 \\
2001 Aug  4 & 0.86 &  9 & 2.59 & 1.2 & 0.09 &  9 & 2.13 & 0.7 & 0.04 \\
2001 Aug 11 & 1.15 & 15 & 2.43 & 1.2 & 0.08 & 15 & 2.14 & 0.9 & 0.06 \\
2001 Sep 20 & 2.12 & 25 & 2.51 & 2.3 & 0.21 & 24 & 2.26 & 3.2 & 0.26 \\
2001 Nov 29 & 1.45 & 20 & 2.88 & 0.7 & 0.07 & 20 & 2.62 & 1.2 & 0.13 \\
2002 Jan  4 & 1.97 & 21 & 2.76 & 0.4 & 0.04 & 21 & 2.52 & 1.6 & 0.12 \\
2002 Feb 21 & 2.07 & 10 & 2.83 & 1.6 & 0.13 & 10 & 2.33 & 1.7 & 0.14 \\
\hline
\multicolumn{10}{@{}l@{}}{\hbox to 0pt{\parbox{170mm}{\footnotesize
\par\indent
\footnotemark[$*$], \footnotemark[$\dagger$] We expect to see rms variations
up to 0.5\% as a result of measurement uncertainties.
}\hss}}
\end{tabular}
\end{center}
\end{table*}
At each epoch, the time spanned by the observations of PKS 1622$-$297 is
given, together with the number of measurements ($N$; usually the same at
4.8 and 8.6\,GHz), the mean flux density ($\bar{S}$), the modulation index
($m = 100 \sigma_{S} / \bar{S}$), and the range of the measured flux
densities ($S_{\mathrm{max}} - S_{\mathrm{min}}$), at each frequency.
The results are shown graphically in figure~\ref{fig:Figure5}.
The upper and lower panels of figure~\ref{fig:Figure5} show the observation
on 2001 June 2 -- 3, when the largest amplitude IDV was observed
($S_{\mathrm{max}} - S_{\mathrm{min}} \sim 0.3$\,Jy), and long-term flux
variation, respectively.
Variations at a level exceeding that expected from measurement uncertainties
are seen at both frequencies at all epochs, with formally significant evidence
of IDV present at at least one frequency at almost all epochs,
whereas \citet{Kedziora01} had succeeded in detecting IDV at only 8.6\,GHz.
PKS 1622$-$297 had the highest mean linear polarization at 4.8\,GHz of the
21 sources in the sample, 5.3\%, with similar percentage linear polarization
at 8.6\,GHz.
The on-axis and linear polarization feed design of the ATCA also enables
accurate measurements of circular polarization \citep{Rayner00}, and
we note significant levels of circularly polarized radio emission were
detected from PKS 1622$-$297.
At 4.8\,GHz, a circularly polarized flux density, averaged over three epochs,
of $-4.8$\,mJy was detected, corresponding to a fractional circular
polarization of $-0.16$\% (the negative sign indicating left-handed
polarization).

A number of models have been proposed for the origin of circular polarization,
ranging from intrinsic processes to propagation effects (see, e.g.,
\cite{Rayner00}).
Given the detection of IDV in PKS 1622$-$297, the model for
scintillation-induced circular polarization is of particular interest
\citep{Macquart00}.
Significant IDV was also observed for PKS 1622$-$297 in Stokes $Q$ and $U$,
however we limit our discussion here to the IDV in the total flux density.



\section{Discussion}
\label{sec:Discussion}

\subsection{Estimation of Brightness Temperature, Doppler Factor, and Jet
Viewing Angle}
\label{subsec:Estimation of Parameters}

From the model fitting parameters we can calculate the source brightness
temperature in the source's rest frame, $T_{\mathrm{B}}$, as
\begin{equation}
T_{\mathrm{B}} = 1.41 \times 10^9 (1+z) (\theta_{\mathrm{maj}}
\theta_{\mathrm{min}})^{-1} S \lambda^2 \;\;\; {\rm K},
\label{eqn:Brightness Temperature}
\end{equation}
where $\theta_{\mathrm{maj}}$ and $\theta_{\mathrm{min}}$ are the FWHM sizes
of the component in the major and minor axes in mas, $S$ is the flux density
of the Gaussian component in Jy, and $\lambda$ is the observation wavelength
in centimeters.
The brightness temperature of each component is given in
table~\ref{tbl:Table1}.
The brightest component C has
$T_{\mathrm{B}} = (3.73 \pm 0.07) \times 10^{11}$\,K.

\begin{figure}
\begin{center}
\FigureFile(80mm,80mm){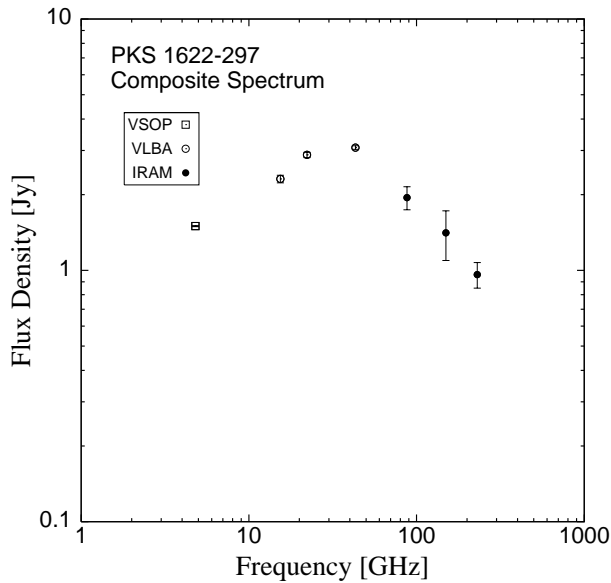}
\end{center}
\caption{Composite radio spectrum of PKS 1622$-$297. Data are from
\citet{Jorstad01a} (VLBA), \citet{Reuter97} (IRAM), and this work (VSOP).
The figure refers only to the flux density of the core component at each
frequency. The data are not obtained simultaneously.}
\label{fig:Figure6}
\end{figure}

The upper limit of the observed brightness temperature is caused by
inverse Compton catastrophe \citep{Kellermann69}.
Using formulae (1a) and (1b) in \citet{Readhead94}, the inverse Compton
scattering limit can be calculated as
$T_{\mathrm{B,ic}} = 1.02 \times 10^{11}$\,K.
To derive this, we adopt a synchrotron peak frequency, $\nu_{\mathrm{peak}}$,
of 43\,GHz, a spectral index of $-0.7$ and assume a synchrotron high-frequency
cutoff of $10^4$\,GHz.
\citet{Jorstad01a} report that the component A, assuming as a core, has an
inverted spectrum between 15.4, 22.2 and 43.2\,GHz from the simultaneous
VLBA observations.
On the other hand, 88, 150, and 230\,GHz single-dish measurements at the
IRAM 30\,m telescope \citep{Steppe93,Reuter97} show that
the source has optically thin spectrum at these frequencies and the
flux densities are lower than that at 43\,GHz by VLBA (see
figure~\ref{fig:Figure6}).
We compiled these results and obtained $\nu_{\mathrm{peak}}$ and $\alpha$,
although each observation was made at a different epoch.
In addition, we adopt a similar value of the synchrotron high-frequency
cutoff obtained in the gamma-ray loud quasar 3C 279 \citep{Piner03}.
To reconcile with our $T_{\mathrm{B}}$, a Doppler factor,
$\delta_{\mathrm{ic}}$, of $2.45 \pm 0.05$ is required.
Here a factor of 0.67 has been used to convert the value of $T_{\mathrm{B}}$,
derived assuming a Gaussian component, to that for an optically thin
uniform sphere \citep{Marscher87}.
The equipartition brightness temperature limit, $T_{\mathrm{B,eq}}$ can
also be estimated assuming the particles and magnetic field are in
equipartition \citep{Readhead94}.
In PKS 1622$-$297, $T_{\mathrm{B,eq}} = 1.03 \times
10^{11}$\,$h^{-0.12}$\,$\delta_{\mathrm{eq}}^{0.84}$\,K.
From our observation we derive $\delta_{\mathrm{eq}}$ as
$(2.87 \pm 0.07)$\,$h^{0.14}$.
Although the obtained $\delta_{\mathrm{eq}}$ is slightly higher than
$\delta_{\mathrm{ic}}$, it seems that the core region of PKS 1622$-$297 is
not far from equipartition at this epoch, in contrast
with the results for several other
core-dominated sources from VSOP observations (NRAO 530,
\cite{Bower98}; PKS 1921$-$293, \cite{Shen99}; PKS 1741$-$038,
\cite{Wajima00}).

The viewing angle, $\phi$, can be obtained using $\beta_{\mathrm{app}}$ and
$\delta$ as
\begin{equation}
\phi = \tan^{-1} \frac{2 \beta_{\mathrm{app}}}{\beta_{\mathrm{app}}^2 +
\delta^2 - 1}
\label{eqn:Viewing Angle}
\end{equation}
\citep{Ghisellini93}.
If each jet component has the same $\delta$ as the nucleus we can obtain
$\phi$ of the components by adopting $\delta_{\mathrm{ic}}$, as
$\timeform{23D.3} \pm \timeform{0D.7}$ (for B2+D2+E3),
$\timeform{13D.2} \pm \timeform{0D.6}$ (for J2), and
$\timeform{6D.9} \pm \timeform{1D.1}$ (for J1), assuming $h = 0.75$.

The derived $\delta$ and $\phi$ of each component are rather different
from other gamma-ray loud AGNs.  Especially, $\delta$ and $\phi$ for
PKS 1622$-$297 are very different from PKS 0528+134 ($\delta \gtrsim
15$, $\phi \sim \timeform{4D}$, \cite{Peng01}) and 3C 279 ($\delta
\gtrsim 20$, $\phi \lesssim \timeform{4D}$, \cite{Piner03};
\cite{Jorstad04}), which show extremely high maximum gamma-ray fluxes of
more than 400\,pJy \citep{Hartman99}.
As mentioned in section \ref{sec:Introduction}, inverse Compton scattering
theory (e.g., \cite{Sikora94}) requires a small jet viewing angle for the
detection of gamma-ray emission in AGN.
Previous theoretical studies imply that objects having larger viewing
angles, $\gtrsim$\timeform{10D}, would be less likely identified as
gamma-ray sources (e.g., \cite{Dermer95}; \cite{Arbeiter02}).
We cannot conclusively explain from our observation results that the
gamma-ray emission from this source is caused by inverse Compton scattering
by relativistic electrons in the parsec-scale jet because the $\phi$ and
$\delta$ of each component, especially components B2+D2+E3 and J2, are
incompatible with other sources.
On the other hand, \citet{Jiang98} report three gamma-ray emitting sources
which have larger viewing angles and lower Doppler factors:
0716+714 ($\phi = \timeform{44D.9}$, $\delta = 1.4$),
Mrk 421 ($\phi = \timeform{35D.2}$, $\delta = 1.3$),
and 1219+285 ($\phi = \timeform{24D.8}$, $\delta = 2.1$).
Further VLBI observations are needed for PKS 1622$-$297 and these three
sources, and further consideration as to whether it is possible to detect
significant gamma-ray emission from AGNs having such large viewing angles.

It is apparent in figure~\ref{fig:Figure4} that the large gamma-ray flare
in 1995.48 occurs soon after the (extrapolated) ejection of component
B2+D2+E3 from the core, a coincidence also noted by \citet{Jorstad01a}.
\citet{Romanova97} developed a dynamical model of gamma-ray flares and VLBI
outbursts of AGNs taking the gamma-ray emission by inverse Compton scattering
into account.
They applied their model to PKS 1622$-$297, fitting the calculated light
curve to that of the gamma-ray flare in 1995.48 observed by EGRET
\citep{Mattox97}, and determined $\phi$ to be \timeform{5D.7}.
This is much smaller than $\phi$ of component B2+D2+E3 obtained from
$\beta_{\mathrm{app}}$ and $\delta_{\mathrm{ic}}$.
Acceptance of their model would then suggest the assumption
that each jet component has the same $\delta$ as the nucleus 
is not completely valid. 

Our observation reveals that PKS 1622$-$297 possesses somewhat different
features to other gamma-ray emitting AGNs.
However, the relatively low brightness temperature we measure may reflect
the fact that the source was in a relatively quiescent state at this epoch.
A similar conclusion was drawn by \citet{Nakagawa05} from their VSOP
observations of NRAO 530.
To reveal the details of the inner region of gamma-ray loud AGN, observations
at a mm-wavelength and higher angular resolution VLBI will be essential.
There are a number of future VLBI projects that will have the capability of
mm-wavelength and higher angular resolution observations, such as
the Korean VLBI Network (KVN; \cite{Minh03}) and the next Space VLBI mission
(VSOP-2; \cite{Hirabayashi04}).
The KVN will be furnished with a multichannel receiver system at 22, 43, 86,
and 129\,GHz and it will be able to provide important information about
wideband spectra, especially in the optically thin regime, for each component.
Higher angular resolution VLBI observations by VSOP-2 will give conclusive
results for the relation between the gamma-ray flare and emergence of the
new VLBI jet component at closer region to the central engine.
Especially, VSOP-2 would be able to make observations in conjunction with
the GLAST gamma-ray spacecraft \citep{Burnett02}, which will provide much
improved angular resolution and sensitivity
($< 6 \times 10^{-9}$counts\,cm$^{-2}$\,s$^{-1}$ for a point source)
compared to EGRET.
This will enable the above subject to be clarified.


\subsection{Intraday Variability}

In the ATCA 2001 June 2 observation, rapid flux density variations of
$\sim 12$\% at both 4.8 and 8.6\,GHz were detected within the first eight
hours (see the upper panels of figure~\ref{fig:Figure5}).
If the observed flux density variations are intrinsic to the source, the
variability brightness temperature, $T_{\mathrm{B,var}}$, can be calculated
as $\sim 10^{19}$\,$h^{-2}$\,K.
This would require a Doppler factor at least two orders of magnitude higher
than estimated in section \ref{subsec:Estimation of Parameters} from the
VSOP data for the intrinsic brightness temperature to remain below the
inverse Compton scattering limit, implying an intrinsic origin for the
IDV is highly unlikely.

As mentioned in section \ref{sec:Introduction}, current radio observations
show that ISS is mainly responsible for the rapid IDV features.
We use the NE2001 Galactic electron density model \citep{Cordes02}
to investigate the likely effect of ISS.
The model predicts a transition frequency between weak and strong scattering
of 31.7\,GHz on the line of sight to PKS 1622$-$297 (Galactic coordinates
$l = \timeform{348D.8}$, $b = \timeform{13D.3}$).
In this case the observed frequencies would be in the strong scattering
regime.
The very rapid IDV observed in a few sources has been shown to be due to
weak scattering at frequencies $\gtrsim 5$\,GHz, in very nearby turbulence
\citep{Dennett00,Rickett02,Bignall03b}.
However the IDV seen in PKS 1622$-$297 has a characteristic timescale
longer than a day.
Our two-day ATCA observations do not provide a sufficient sample of the
variability pattern to estimate the characteristic timescale accurately.
In many epochs we observe only a steady increase or decrease in flux density.
Even for the most extreme variations observed, in 2001 June as shown in
figure~\ref{fig:Figure5}, the variability pattern is not well-sampled.
A characteristic timescale of several days is consistent with refractive
interstellar scintillation (RISS) in the strong scattering regime as would
be expected at 4.8 and 8.6\,GHz for a source along this line of sight
through the Galaxy.

If the transition frequency is close to 30\,GHz, as expected from the NE2001
model, then applying the theory of strong scattering \citep{Walker98}
we would expect to see rms modulations of $m \approx 35$\% at 4.8\,GHz
and $m \approx 50$\% at 8.6\,GHz for a source of angular size
$\theta_{\mathrm{s}}$ smaller than the scattering disk
$\theta_{\mathrm{ref}}$.
If $\theta_{\mathrm{s}} > \theta_{\mathrm{ref}}$, then the modulation index
will be reduced and the scintillation timescale increased by a factor
$\sim \theta_{\mathrm{s}} / \theta_{\mathrm{ref}}$.
The NE2001 model predicts scattering disk sizes of 0.16\,mas at 4.8\,GHz
and 0.04\,mas at 8.6\,GHz for the line of sight to PKS 1622$-$297.
The fitted angular size of component C from the 5\,GHz VSOP data is then
3--4 times larger than the predicted scattering disk, we could therefore
expect to see rms modulations of order 10\% or $\sim 150$\,mJy in this
component as a result of RISS.
The expected characteristic timescale would be of order one week, depending
on the transverse bulk velocity of the scattering plasma with respect to
the observer.
It may also be the case that some fraction of the flux density of component
C is in an even more compact component that is unresolved on VLBI scales,
so that this would scintillate on a shorter timescale than the more
extended component.
After taking into account that PKS 1622$-$297 has significant flux density,
$\sim 1$\,Jy, which is extended on mas to arcsecond scales, the rms
variations observed in the ATCA data remain somewhat lower in amplitude than
expected from the RISS model.
However, since we do not fully sample the scintillation pattern in the
ATCA observations, the modulation index is almost certainly underestimated.
We conclude that the IDV observed in ATCA data is consistent with what is
expected due to RISS of the core component in PKS 1622$-$297.

As RISS is a stochastic process, sampling of the variability over many
more days would be required to accurately estimate the rms flux density
variations and characteristic timescales.
However, on longer timescales this source is also likely to show intrinsic
variability which may complicate analysis.

With the addition of PKS 1622$-$297, short timescale variability within
a few hours to a few days at cm-wavelengths has been reported in 28 sources
(42\%) of 67 EGRET-identified AGNs to date (\cite{Quirrenbach92};
\cite{Wagner95}; \cite{Kedziora01}; and references therein).
This compares with a general incidence for IDV among flat-spectrum radio
sources of between 10 and 20\% (e.g., \cite{Kedziora01}; \cite{Lovell03}),
suggesting a higher incidence of IDV among gamma-ray sources, a possibility
which has been noted previously \citep{Wagner95}.
If the IDV is due to ISS, each source should have a compact core between
several tens to a few hundred $\mu$as, corresponding to a linear scale
of about $10^{17}$\,cm, depending on the source's distance.
This linear scale of the core is similar to, or slightly larger than, those
obtained by \citet{Sikora97}, assuming an external radiation Compton model
(e.g. \cite{Sikora94}) can account for the gamma-ray emission.
Again, the VSOP-2 mission will enable the inner region of the core to be
observed with a maximum angular resolution of 38\,$\mu$as, comparable
to the linear scale of the gamma-ray emission region.
VSOP-2 would certainly be able to tell us more about how the 
core flux density is
distributed and would allow us to measure scatter-broadening in sources
which are moderately strongly scattered.



\section{Conclusions}
\label{sec:Conclusions}

We have observed the gamma-ray loud quasar PKS 1622$-$297 using the HALCA
satellite and three ground radio telescopes.
The source was clearly resolved and we could determine three distinct
features, the core and two jet components, and have confirmed that both
jet components display superluminal motion.
By using the inverse Compton catastrophe model we estimated the Doppler
factor, $\delta$, as 2.45, and the jet viewing angle, $\phi$, of each
component as larger than \timeform{6D.9}.
Especially, the innermost component has very large viewing angle,
$\phi = \timeform{23D.3}$.
Both $\delta$ and $\phi$ are rather different from other gamma-ray loud
AGNs yet observed, though this may be due to the quasar being in
a relatively quiescent (i.e., low brightness temperature) state
during the epoch of the VSOP observation.
Although some previous theoretical and observational studies have implied
the possibility of inverse Compton scattering as the gamma-ray emission
mechanism, the VSOP observations at epoch 1998.15 do not on their own
provide any supporting evidence.

We have also presented the results of the ATCA total flux density
monitoring between February 2001 and February 2002. 
PKS 1622$-$297 displays IDV at both 4.8 and 8.6\,GHz throughout this year.
The extremely high brightness temperature implied for the IDV being
due to fluctuations intrinsic to the source strongly favor refractive
interstellar scintillation as the origin of the IDV.

In either case, compact structure at a level significantly below the
angular resolution of our VSOP observations can be inferred.
Scattering models imply the existence of a compact core, $\sim 0.1$\,mas
in angular size, or $\sim 10^{17}$\,cm in linear size.
This is comparable to the gamma-ray emitting region expected from the
theoretical model.

\bigskip

We are grateful to the referee, Dr.\ David L.\ Jauncey, for valuable comments
which improved the manuscript.
We gratefully acknowledge the VSOP Project, which is led by the Japanese
Institute of Space and Astronautical Science of the Japan Aerospace
Exploration Agency in cooperation with many organizations and radio
telescopes around the world.
Especially, we acknowledge three ground telescopes, Hartebeesthoek, Hobart,
and Shanghai, for participating the VSOP observation.
The research was carried out using an observation of the VSOP Survey Program.
The research has made use of data from the University of Michigan Radio
Astronomy Observatory which has been supported by the University of Michigan
and the National Science Foundation.
The Australia Telescope Compact Array is part of the Australia Telescope
which is funded by the Commonwealth of Australia for operation as a
National Facility managed by CSIRO.
KW acknowledges support from Korea Science and Engineering Foundation
(KOSEF).





\begin{thebibliography}{}

\bibitem[Arbeiter et~al.(2002)]{Arbeiter02}
Arbeiter,~C., Pohl,~M., \& Schlickeiser,~R.\ 2002, \aap, 386, 415

\bibitem[Balbi et~al.(2000)]{Balbi00}
Balbi,~A., \etal\ 2000, \apj, 545, L1 (erratum 2001, 558, L145)

\bibitem[Bignall(2003a)]{Bignall03a}
Bignall,~H.~E.\ 2003a, PhD Thesis, University of Adelaide

\bibitem[Bignall et~al.(2003b)]{Bignall03b}
Bignall,~H.~E., \etal\ 2003b, \apj, 585, 653

\bibitem[Bower, Backer(1998)]{Bower98}
Bower,~G.~C., \& Backer,~D.~C.\ 1998, \apj, 507, L117

\bibitem[Burnett et~al.(2002)]{Burnett02}
Burnett,~T.~H., \& GLAST Team 2002, \baas, 34, 1315

\bibitem[Carlson et~al.(1999)]{Carlson99}
Carlson,~B.~R., Dewdney,~P.~E., Burgess,~T.~A., Casorso,~R.~V.,
Petrachenko,~W.~T., \& Cannon,~W.~H.\ 1999, \pasp, 111, 1025

\bibitem[Cordes, Lazio(2002)]{Cordes02}
Cordes,~J.~M., \& Lazio,~T.~J.~W.\ 2002, astro-ph/0207156

\bibitem[Dennett-Thorpe, de~Bruyn(2000)]{Dennett00}
Dennett-Thorpe,~J., \& de~Bruyn,~A.~G.\ 2000, \apj, 529, L65

\bibitem[Dermer(1995)]{Dermer95}
Dermer,~C.~D.\ 1995, \apj, 446, L63

\bibitem[Fomalont et~al.(2000)]{Fomalont00}
Fomalont,~E.~B., Frey,~S., Paragi,~Z., Gurvits,~L.~I., Scott,~W.~K.,
Taylor,~A.~R., Edwards,~P.~G., \& Hirabayashi,~H.\ 2000, \apjs, 131, 95

\bibitem[Ghisellini et~al.(1993)]{Ghisellini93}
Ghisellini,~G., Padovani,~P., Celotti,~A., \& Maraschi,~L.\ 1993,
\apj, 407, 65

\bibitem[Hartman et~al.(1999)]{Hartman99}
Hartman,~R.~C., \etal\ 1999, \apjs, 123, 79

\bibitem[Hirabayashi et~al.(1998)]{Hirabayashi98}
Hirabayashi.~H., \etal\ 1998, Science, 281, 1825 (erratum 1998, 282, 1995)

\bibitem[Hirabayashi et~al.(2000)]{Hirabayashi00}
Hirabayashi.~H., \etal\ 2000, \pasj, 52, 997

\bibitem[Hirabayashi et~al.(2004)]{Hirabayashi04}
Hirabayashi.~H., \etal\ 2004, \procspie, 5487, 1646

\bibitem[Impey, Tapia(1990)]{Impey90}
Impey,~C.~D., \& Tapia,~S.\ 1990, \apj, 354, 124

\bibitem[Jackson et~al.(2002)]{Jackson02} 
Jackson,~C.~A., Wall,~J.~V., Shaver,~P.~A., Kellermann,~K.~I., 
Hook,~I.~M., \& Hawkins,~M.~R.~S.\ 2002, \aap, 386, 97 

\bibitem[Jauncey et~al.(2003)]{Jauncey03}
Jauncey,~D.,~L., Bignall,~H.~E., Lovell,~J.~E.~J., Kedziora-Chudczer,~L.,
Tzioumis,~A.~K., Macquart,~J.~-P., \& Rickett,~B.~J. 2003, in ASP Conf.\
Ser.\ 300, Radio Astronomy at the Fringe, ed.\ J.~A.~Zensus, M.~H.~Cohen,
\& E.~Ros (San Francisco: ASP), 199

\bibitem[Jauncey, Macquart(2001)]{Jauncey01}
Jauncey,~D.,~L., \& Macquart,~J.~-P. 2001, \aap, 370, L9

\bibitem[Jiang, Cao, and Hong(1998)]{Jiang98}
Jiang,~D.~R., Cao,~X., \& Hong,~X.\ 1998, \apj, 494, 139

\bibitem[Jorstad et~al.(2004)]{Jorstad04}
Jorstad,~S.~G., Marscher,~A.~P., Lister,~M.~L., Stirling,~A.~M.,
Cawthorne,~T.~V., G\'omez,~J.~-L., \& Gear,~W.~K.\ 2004, \aj, 127, 3115

\bibitem[Jorstad et~al.(2001b)]{Jorstad01b}
Jorstad,~S.~G., Marscher,~A.~P., Mattox,~J.~R., Aller,~M.~F., Aller,~H.~D.,
Wehrle,~A.~E., \& Bloom,~S.~D.\ 2001b, \apj, 556, 738

\bibitem[Jorstad et~al.(2001a)]{Jorstad01a}
Jorstad,~S.~G., Marscher,~A.~P., Mattox,~J.~R., Wehrle,~A.~E., Bloom,~S.~D.,
\& Yurchenko,~A.~V.\ 2001a, \apjs, 134, 181

\bibitem[Kedziora-Chudczer et~al.(2001)]{Kedziora01}
Kedziora-Chudczer,~L.~L., Jauncey,~D.~L., Wieringa,~M.~H., Tzioumis,~A.~K.,
\& Reynolds,~J.~E.\ 2001, \mnras, 325, 1411

\bibitem[Kellermann, Pauliny-Toth(1969)]{Kellermann69}
Kellermann,~K.~I., \& Pauliny-Toth,~I.~I.~K.\ 1969, \apj, 155, L71

\bibitem[K\"onigl(1981)]{Konigl81}
K\"onigl,~A.\ 1981, \apj, 243, 700

\bibitem[Lovell et~al.(2003)]{Lovell03}
Lovell,~J.~E.~J., Jauncey,~D.~L., Bignall,~H.~E., Kedziora-Chudczer,~L.,
Macquart,~J.~-P., Rickett,~B.~J., \& Tzioumis,~A.~K.\ 2003, \aj, 126, 1699 

\bibitem[Macquart, Melrose(2000)]{Macquart00}
Macquart,~J.~-P, \& Melrose,~D.~B.\ 2000, \apj, 545, 798

\bibitem[Marscher(1987)]{Marscher87}
Marscher,~A.~P.\ 1987, in Superluminal Radio Sources, ed.\ J.~A.~Zensus \&
T.~J.~Pearson (Cambridge and New York: Cambridge University Press), 280

\bibitem[Mattox et~al.(1997)]{Mattox97}
Mattox,~J.~R., Wagner,~S.~J., Malkan,~M., McGlynn,~T.~A., Schachter,~J.~F.,
Grove,~J.~E., Johnson,~W.~N., \& Kurfess,~J.~D.\ 1997, \apj, 476, 692

\bibitem[Minh et~al.(2003)]{Minh03}
Minh,~Y.~C., Roh,~D.~-G., Han,~S.~-T., \& Kim,~H.~-G.\ 2003, in ASP Conf.\
Ser.\ 306, New Technologies in VLBI, ed.\ Y.~C.~Minh (San Francisco: ASP), 373

\bibitem[Nakagawa et~al.(2005)]{Nakagawa05}
Nakagawa,~A., Edwards,~P.~G., Murata,~Y., Wajima,~K., \& Omodaka,~T.\ 2005,
\pasj, 57, 295 

\bibitem[Peng et~al.(2001)]{Peng01}
Peng,~B., \etal\ 2001, \apj, 551, 172

\bibitem[Piner et~al.(2003)]{Piner03}
Piner,~B.~G., Unwin,~S.~C., Wehrle,~A.~E., Zook,~A.~C., Urry,~C.~M., \&
Gilmore,~D.~M.\ 2003, \apj, 588, 716

\bibitem[Quirrenbach et~al.(1992)]{Quirrenbach92}
Quirrenbach,~A., \etal\ 1992, \aap, 258, 279

\bibitem[Rayner et~al.(2000)]{Rayner00}
Rayner,~D.~P., Norris,~R.~P., \& Sault,~R.~J.\ 2000, \mnras, 319, 484

\bibitem[Readhead(1994)]{Readhead94}
Readhead,~A.~C.~S.\ 1994, \apj, 426, 51

\bibitem[Reuter et~al.(1997)]{Reuter97}
Reuter,~H.~-P., \etal\ 1997, \aaps, 122, 271

\bibitem[Rickett et~al.(2001)]{Rickett01}
Rickett,~B.~J., Witzel,~A., Kraus,~A., Krichbaum,~T.~P., \& Qian,~S.~J.
2001, \apj, 550, L11

\bibitem[Rickett et~al.(2002)]{Rickett02}
Rickett,~B.~J., Kedziora-Chudczer,~L., \& Jauncey,~D.~L.\ 2002, \apj, 581, 103

\bibitem[Romanova and Lovelace(1997)]{Romanova97}
Romanova,~M.~M., \& Lovelace,~R.~V.~E.\ 1997, \apj, 475, 97

\bibitem[Saikia et~al.(1987)]{Saikia87}
Saikia,~D.~J., Singal,~A.~K., \& Cornwell,~T.~J.\ 1987, \mnras, 224, 379

\bibitem[Sault et~al.(1995)]{Sault95}
Sault,~R.~J., Teuben,~P.~J., \& Wright,~M.~C.~H.\ 1995, in ASP Conf.\ Ser.\
77, Astronomical Data Analysis Software and Systems IV, ed.\ R.~A.~Shaw,
H.~E.~Payne, \& J.~J.~E.~Hayes (San Francisco: ASP), 433

\bibitem[Shen et~al.(1999)]{Shen99}
Shen,~Z.~-Q., Edwards,~P.~G., Lovell,~J.~E.~J., Fujisawa,~K., Kameno,~S.,
\& Inoue,~M.\ 1999, \pasj, 51, 513

\bibitem[Shepherd(1997)]{Shepherd97}
Shepherd,~M.~C.\ 1997, in ASP Conf.\ Ser.\ 127, Astronomical Data Analysis
Software and Systems VI, ed.\ G.~Hunt \& H.~E.~Payne (San Francisco: ASP), 77

\bibitem[Sikora et~al.(1994)]{Sikora94}
Sikora,~M., Begelman,~M.~C., \& Rees,~M.~J.\ 1994, \apj, 421, 153

\bibitem[Sikora et~al.(1997)]{Sikora97}
Sikora,~M., Madejski,~G., Moderski,~R., \& Poutanen,~J.\ 1997, \apj, 484, 108

\bibitem[Sowards-Emmerd et~al.(2003)]{Sowards03}
Sowards-Emmerd,~D., Romani,~R.~W., \& Michelson,~P.~F.\ 2003, \apj, 590, 109 

\bibitem[Sowards-Emmerd et al.(2004)]{Sowards04}
Sowards-Emmerd,~D., Romani,~R.~W., Michelson,~P.~F., \& Ulvestad,~J.~S.\ 2004,
\apj, 609, 564 

\bibitem[Steppe et~al.(1993)]{Steppe93}
Steppe,~H., \etal\ 1993, \aaps, 102, 611

\bibitem[Stickel et~al.(1994)]{Stickel94}
Stickel,~M., Meisenheimer,~K., \& K\"uhr,~H.\ 1994, \aaps, 105, 211

\bibitem[Tabara, Inoue(1980)]{Tabara80}
Tabara,~H., \& Inoue,~M.\ 1980, \aaps, 39, 379

\bibitem[Tingay et~al.(2003)]{Tingay03} 
Tingay,~S.~J., Jauncey,~D.~L., King,~E.~A., Tzioumis,~A.~K., 
Lovell,~J.~E.~J., \& Edwards, P.~G.\ 2003, \pasj, 55, 351

\bibitem[Tingay, Murphy, and Edwards(1998)]{Tingay98}
Tingay,~S.~J., Murphy,~D.~W., \& Edwards,~P.~G.\ 1998, \apj, 500, 673

\bibitem[Tingay et~al.(2002)]{Tingay02}
Tingay,~S.~J., \etal\ 2002, \apjs, 141, 311

\bibitem[Wagner, Witzel(1995)]{Wagner95}
Wagner,~S.~J., \& Witzel,~A.\ 1995, \araa, 33, 163

\bibitem[Wajima et~al.(2000)]{Wajima00}
Wajima,~K., Lovell,~J.~E.~J., Kobayashi,~H., Hirabayashi,~H., Fujisawa,~K.,
\& Tsuboi,~M.\ 2000, \pasj, 52, 329

\bibitem[Walker(1998)]{Walker98}
Walker,~M.~A.\ 1998, \mnras,  294, 307

\bibitem[Wright and Otrupcek(1990)]{Wright90}
Wright,~A., \& Otrupcek,~R.\ 1990, Parkes Catalogue, Australia Telescope
National Facility, CSIRO

\end{thebibliography}
\end{document}